# Science Education in the 21st Century

## Sun Kwok[1]

Sun Kwok was Dean of Science of the University of Hong Kong from 2006 to 2016.

Email: sunkwok@hku.hk

Should science be taught differently? By emphasizing the process, not factual knowledge, students will learn how to approach and solve problems and see science as relevant to their careers outside of research.

## Abstract

The traditional university science curriculum was designed to train specialists in specific disciplines. However, in universities all over the world, science students are going into increasingly diverse careers and the current model does not fit their needs. Advances in technology also make certain modes of learning obsolete. In the last 10 years, the Faculty of Science of the University of Hong Kong has undertaken major curriculum reforms.  A sequence of science foundation courses required of all incoming science students are designed to teach science in an integrated manner, and to emphasize the concepts and utilities, not computational techniques, of mathematics. A number of non-discipline specific common core courses have been developed to broaden students' awareness of the relevance of science to society and the interdisciplinary nature of science. By putting the emphasis on the scientific process rather than the outcome, students are taught how to identify, formulate, and solve diverse problems.

The purpose of institutions of higher learning has evolved over the past millennium, from praising the glory of God, to self-fulfillment, to the search for truth. The 20th century saw the gradual introduction of agriculture and mining schools, teacher colleges, and business schools to serve society's practical needs. Most universities now use a mixed model. In addition to core subjects such as arts and science, colleges generally include professional disciplines such as architecture, business, engineering, law, and medicine.

---

[1] Current address: Department of Earth, Ocean & Atmospheric Sciences, University of British Columbia, Vancouver, B.C., Canada V6T 1Z4

So what is the purpose of universities in the 21st century?  Should they be preparatory schools that provide professional or vocational training?  Many educators maintain that a university education should develop students as people, prepare them to think, and set the foundation for life-long self-learning and self-improvement.  Unfortunately, there is often a mismatch between educators' and students' expectations.  Many students believe universities will provide them with a meal ticket for a better job.  However, universities are not diploma mills; students should attend not just to get a degree, but to learn how to evolve their own intellect.

What are the problems with current science education?

Science education at all levels of schooling is often seen as abstract and irrelevant to real life.  Students in chemistry and biology are burdened with memorization of facts, and students in physics and mathematics feel that their discipline contents are abstract and cannot relate these materials to the real world.  Students in general fail to see that science is in Nature all around them, and the scientific method is widely applicable in different aspects of their lives.

In school curricula in most parts of the world, science subjects are segregated into physics, chemistry, and biology, and the connections between these fields are usually not emphasized.  Memorization, rote-learning, and keyword marking in exams are well-known ills.  Young people regard schooling as a game to be won but not a road to intellectual fulfillment.

While many of my colleagues in North America lament their students' lack of technical abilities, I note that Asian students can calculate very quickly and accurately but have no idea that mathematics is an essential framework for science.  They are able to derive complex equations in physics, but they fail to relate these equations to physics in Nature.  To them, it is just an academic exercise.  So the problem is not just how much science students learn but how they connect science to their lives and society.  Although many Asian universities turned out students with very impressive calculation skills, their ability to apply their learnings to society and innovate were not previously nurtured.

We are also faced with rapid technological change. A lot of factual information is available on the internet, and artificial intelligence is making certain occupations obsolete.  It is therefore much more important to give our students fundamentals that will stay with them for the rest of their lives.  These essential tools include language skills such as comprehension, expression, and communication, as well as quantitative skills such as analysis, seeing hidden patterns, identifying variables, and formulating solutions to problems.

The current landscape

There are significant variations in how science is taught at the undergraduate level.  Under the Soviet (Russia and China) and British systems, students receive a focused, specialized education.  Since Hong Kong was a British Colony, The University of Hong Kong (HKU) followed the British tradition.  Students used to apply to a specific program (e.g., biochemistry) for university admission. Upon entry, they follow a fixed recipe of courses, almost entirely in a single discipline.

In comparison, the United States has a liberal education philosophy centered around a designated major, distribution requirements, and electives. Although science students are often required to take courses in humanities and social sciences, these courses are usually just introductory courses in their respective disciplines and not specifically designed for science students.

At research-intensive universities, the science curricula are often designed to train students to be the next generation of academics. But only a small fraction of graduates get PhDs and become professors. Many science graduates become leaders in government, NGOs, and the private sector. Our science training should help them fulfill these roles.

Rationale and objectives for science education reform

What is the purpose of university science education? It should not be just to train scientists but to introduce students to a scientific way to thinking that will make them better citizens[1]. Science education benefits not only the individual but society as a whole. In a democratic country, the collective views of citizens influence the nation's directions[2]. The inability of common people to understand and interpret graphs, statistics, and scientific data could threaten democracy.

While current science curricula often focus on factual knowledge content, I argue that it is more important to teach the process of science. Training should include mastering methods such as building models, constructing experiments, taking data, revising models based on data, and communicating results. Students should acquire the ability to solve problems by studying examples of previous work. In the process, they should develop free, bold, independent, and creative thinking. They should be able to make rational judgments and rise above the ignorance and prejudice that are prevalent in society.

Science students should be encouraged to develop their sense of curiosity and acquire the confidence to ask questions and challenge assumptions. Science students should be knowledgeable about our world and aware of how Nature works. They should also think analytically and quantitatively, keep an open mind, and remain independent from public opinion. Our goal is to train students as people of intellect, not for a vocation. Graduates should be versatile enough to take on any job. Most importantly, their education should lay the ground work for lifelong learning, as society's needs are constantly changing.

In 2005, the University Grants Council of Hong Kong initiated a general reform of higher education, to be implemented in 2012[3]. As the first executive Dean of Science after the HKU governance reform, I was given the mandate in 2006 to overhaul the science curriculum. My goal was to create a broader education with a flexible curriculum. We learned from the North American system but tried to go further. While there is a large volume of literature on theories of science education, we had the unique opportunity of actually putting some of these theories into practice.

Our curriculum reform was designed with the following objectives:

- Prepare students as holistic individuals who can think analytically, solve a diverse set of problems, and communicate the results;
- Provide opportunities for students to interact with our community of scholars and engage in research and practical exercises;
- Expose students to diverse academic training and practical experiences beyond the classroom and laboratory;
- Provide a broad academic background in all the sciences.

Reactions to the reform and the Hong Kong paradox

When I first suggested reforming the science curriculum, I faced some skeptics who pointed out that: "Hong Kong was already doing very well in PISA (Programme for International Student Assessment) results. In 2012, we were third in the world in math, and second in science and reading. Why did we need to change? My answer was that our students may do well in exams, but they don't know how science and mathematics are relevant to their lives. My own interactions with students suggest that while they can master the techniques, they often don't know what they are for. They may be the best calculators in the world, but they are unable to apply these techniques to real-world problems.

There are also dissenting opinions on the broader issue of education philosophy. These critics can be classified into two groups. The first group (mostly from the business sector) maintains that as long as we admit the best students, they will excel no matter what we do. Our students make personal connections in the residential halls, and networking is all they need to succeed in the business world.

The second group believes that university graduates need to learn technical skills and be useful as "little screws" in machinery. They don't need a liberal education. In fact, thinking makes trouble makers. Students should focus on their subject major.

Since I am a scientist and educator, I do not agree with these two views. But I have to admit that there are different philosophies of education, reflecting different economic and political viewpoints.

Science curriculum reform at HKU

We undertook a multi-year reform. Some steps were taken to catch up with existing practices in North America, but some were original innovations. These reforms are listed below by date of implementation.

- 2006: Change from the highly specialized program-based curriculum to a credit-based major/minor system.
- 2007: Change from discipline-based admission to faculty common admission. Students are free to choose from a large number of majors.
- 2007: Compulsory experiential learning for all students.

- 2007: Student advisory system.  Since students have more freedom and choices, they need more advice and guidance.
- 2008: Academic induction.  Two weeks before the first term, freshmen are enrolled in academic induction to learn about the differences between learning in university and in high school.
- 2010: Common Core courses introduced by the University.
- 2012: Science Foundation courses implemented in the Faculty of Science

The last two items are the most innovative parts of our reform and I will discuss them in detail below.

Science foundation courses

A key component of our science curriculum reform was the introduction of Science Foundation courses in 2012.  This sequence of two 1-semester courses is required of all incoming science students, irrespective of their intended major.  These courses are a departure from the previous segregated way of learning science.  The goal is to give students a broad view of science's nature, history, fundamental concepts, methodology, and impact on civilization and society. The classes are also intended to help students make informed decisions when selecting a major.

The first course focuses on the scientific method and reasoning.  It teaches the nature, history, and methodology of science and is intended to equip students with basic logical and quantitative reasoning skills.  The course surveys mathematics and statistics and explores their applications.  Specifically, we discuss what mathematics is, the various types of mathematics, and how mathematics aid different disciplines.  The emphasis is on the formulation of problems but not calculations.  Some examples of course content include the scientific method, hypotheses and logical deduction, observation and experiments, inductive and deductive reasoning, criteria for good scientific theories, guesstimation, mathematical modelling, difference equations, matrices, differential equations, fractals and chaos, hypothesis testing and decision making using statistics, correlation and causation, regression, and predictions.

The second course is called Fundamentals of Modern Science, and the goal is to provide students with an overview of natural science.  Our course encompasses the traditional subjects of physics, astronomy, earth sciences, chemistry and biology and introduces the general principles and unifying concepts to describe diverse natural phenomena.  We start from the smallest building blocks of matter, moving from elementary particles, to atoms, molecules, cells, living organisms, the Earth, and the Universe, emphasizing the relationships between science subjects.

These Science Foundation courses share some teaching and assessment techniques.  In both courses, we introduce the history of the subject.  We start with real life problems, not abstract concepts. In exams, students are not asked to repeat tasks from lectures but given new situations that they need to have mastered the taught materials to solve.

An example of an interdisciplinary course

The second component of the reform was the Common Core courses, which were introduced by the University in 2010.  After 2012, all HKU students were required to take six common core courses.  These non-discipline-based classes are designed to develop broader perspectives, critical assessment of complex issues, appreciation of our and other cultures, and the qualities necessary to be a member of the global community.

I would like to use a course that I developed in 2010 as an illustrative example.  I taught the course eight times from 2010 to 2017.  Every year, I had approximately 100 students from all faculties of the University, including Arts, Business, Education, Engineering, Law, Medicine, Science, and Social Sciences.  The course "Our Place in the Universe" discussed the changing perceptions of our place in the universe as the result of astronomical development. It illustrated the development of the scientific method and how science has influenced our philosophical thinking and cultural development.

The course used the historical development of astronomy to illustrate the process of rational reasoning and its effect on philosophy, religion, and society. Because celestial objects followed regular patterns, astronomical observations gave humans some of the first hints that Nature was understandable.  The complicated nature of these patterns also challenged our intellectual powers.

The approach of this course was different from a typical science course, where science is presented as a series of facts or abstract concepts.  Instead, I emphasized that science is about the process of rational thinking and creativity.   What we consider to be the truth is constantly evolving and has changed greatly over the history of humankind.  The essence of science is not so much about the current view of our world, but how we changed from one set of views to another.  This course was not about the outcome but the process[4].

I began with a description of basic observations of celestial objects, summarized the patterns observed and the problems they posed, and discussed the suggested theories and their implications.  The pros and cons of these theories were evaluated alongside alternate theories.  In contrast, typical science textbooks take an axiomatic approach by first stating the correct theory and deriving the deductions before comparing them with experimental results.  I hope this historical approach allowed students to better understand the scientific process and learn from this process when they tackle real-life problems in their careers.

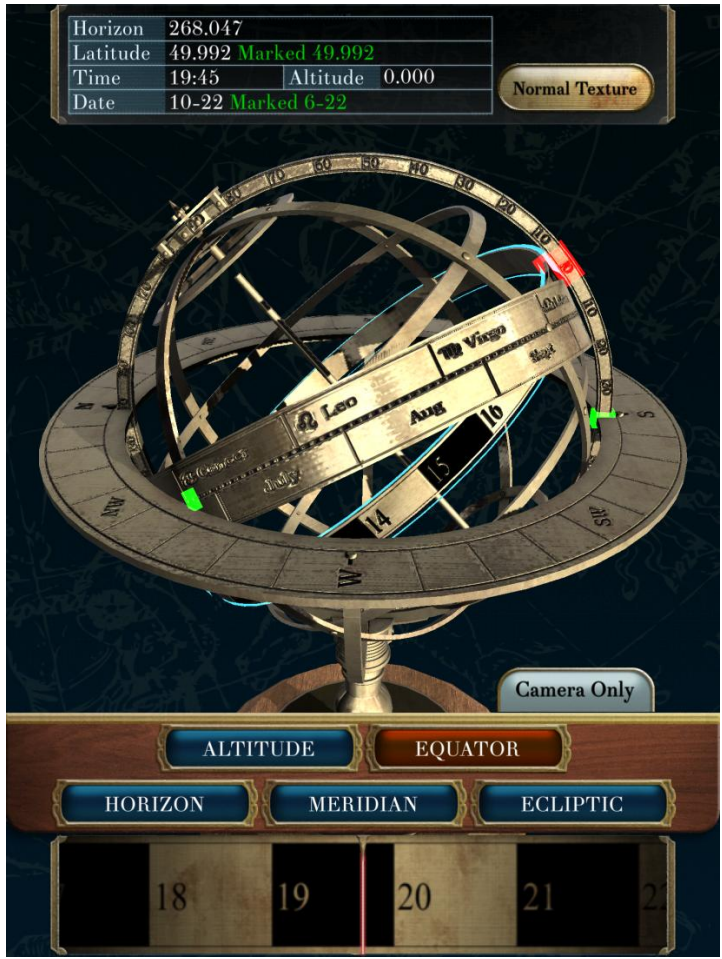

Fig. 1. An armillary sphere mobile app designed to help students understand the power of prediction of the geocentric model. By manipulating this app, students can determine the time and direction of sunrise/sunset, altitude of the Sun at noon, the dates when the Sun can be overhead, at any place on Earth for any date of the year. This app is available free from both Apple and Android platforms. Image credit: HKU Technology-Enriched Learning Initiative.

Training to think, not to recite

Students ask whether they will be handicapped without previous training in physics or astronomy. In fact, science and engineering students have the most difficulty in the course because they have been told all the modern notions but never learned how we arrived at those conclusions.

It has been very illuminating to ask students: "How do we know that the Earth revolves around the Sun?" Every student accepts this as the truth, but if pressed, no one can answer this question other than to say "This is what I was told by my teacher". Science education is supposed to promote an enquiring mind, but what we have been doing is exactly the opposite. Since science teaching is often authoritative, beginning with the correct model and deriving all its consequences, students get the mistaken idea that truth is easy to arrive at and self-evident.

Actually, the apparent motions of the Sun, Moon, stars and planets are regular but complicated. The geocentric model was extremely successful in explaining and predicting these motions. Copernicus' model was more complex (has more epicycles), not more accurate, and was contrary to apparent common sense. The advantages of Copernicus' model were subtle and evidence came much later. Only after students have gone through the evolution of cosmological models do they appreciate the true significance of the Copernican revolution.

My goals in the course were to encourage students to observe their surroundings, rekindle their childhood curiosity, and learn to be aware of natural phenomena. Computer labs were used to help students make simulated observations of the Sun, the Moon and the planets, and a mobile-device app was developed to demonstrate the use of the armillary sphere and the power of the geocentric model (Figure 1). From these observations, they learned how to think independently, logically and analytically. Through historical examples, they learned how to solve problems.

What have they learned?

Astronomy is just one example of the evolution of rational thinking. Answering the question of whether the Earth revolves around the Sun is not easy, and the same is true for concepts in biological evolution and climate change. Our search for truth is long and hard and requires strong perseverance. Copernicus took decades to repeat Ptolemy's calculations in the framework of the heliocentric model and Kepler took many years to figure out that Mars' orbit was an ellipse. This is one of the major messages that students got from the course.

Can this model be applied elsewhere?

As a science course with history and philosophy components, this is a highly unusual course. The goal of this course is not to teach astronomy but to use astronomy to illustrate how humans developed rational thinking. We do discuss astronomical facts, but these are only tools to illustrate that our world is complex and how human ingenuity managed to explain such complexity. The goal is that the skills learned by students in this course are applicable not only in astronomy or science, but in other disciplines as well.

Many North American universities have distribution requirements asking non-science students to take one or more science courses (and vice versa). I believe that a course like "Our Place in the Universe" is more useful for students' intellectual development than what they may gain through an introductory physics, biology, economics, or psychology course. On the other hand, it is much more expensive and time-consuming to develop and run. One possible model that other universities could adopt is to develop a small set of courses that are fundamental to a university education (see, for example, the goals of common core above). Economy of scale can be achieved by offering these courses in multiple sessions. The challenge is to find suitable teachers, as not all teachers are equipped to teach interdisciplinary courses. Of course, if these reforms are adopted on a large scale, then future teachers who graduate from such programs may be better prepared.

Reform of disciplinary courses

In this Internet age, many scientific facts can be found online. Factual details are less important than concepts. Computing devices are commonly used for all mathematical and statistical analysis and less time should be spent on drills to practice the techniques. Although the reform of disciplinary courses at HKU was left to individual departments, I asked each department to consider the following points:

- Are all the courses in our major requirements necessary?
- All disciplinary courses should focus on the fundamentals and not be overly specialized
- Courses should be designed for the benefit of students, not the convenience of the instructors. Some courses are in the syllabus because they reflect the research specialties of individual faculty members, which may not be essential to a particular major
- The major program should be flexible enough to allow the most discipline-dedicated students to gain in-depth knowledge and have a successful career in graduate school but also allow students who do not take this path to have a broad and meaningful education

Many new areas of science are inter-disciplinary, and our curriculum must reflect this. Examples are environmental science, astrobiology, and global change. Our traditional major structures need to be re-examined.

How do we assess the outcome?

How do we evaluate the success of these reforms? The most common approach is to rely on student evaluations. But students' goals may not always align with our educational objectives. Students are often concerned about workload and grades, and they may not immediately appreciate the educational benefits. Student evaluations are good at gauging student satisfaction but do not accurately measure the effectiveness of learning[5]. While student feedback is essential and valuable, we must carefully word the questions interpret the results. An accurate evaluation of the success of curriculum reform is not easy, and will require long-term engagements with students, such as follow-up survey several years after graduation. Specifically, we would like to learn from these surveys whether the reforms have improved students' intellectual development and enabled them to be better prepared in entering a diverse range of careers.

Challenges to science education reform

Recent teaching reforms in North America have emphasized changes in teaching methods[6]. The goal is to deliver better results than the traditional lecture-tutorial format. Many institutions are trying interactive learning, flipped classrooms, group learning, student engagement, and on-line learning. However, curriculum changes have been minor in comparison. I have given talks on our curriculum reform at universities in Asia and North America, and the reactions are almost identical. Many senior administrators such as provosts and deans recognize the inadequacies of current science curricula, and they would like their curriculum to benefit students who pursue diverse careers. At the same time, they say that actual reforms are difficult as curricula in universities are controlled by academic departments, faculty/college councils, and university senates, all dominated by academic staff who may not share this goal.  Many academics consider it is their duty to maintain high standards in their discipline in order to prepare their students for a research career, and they believe that any alternate mode of education will dilute their training.

The first step to curriculum reform is therefore to prepare a comprehensive package of reform and to persuade our colleagues to vote for such a reform.   The second step is to find colleagues who are able to teach the new kinds of courses (the science foundation and common core courses mentioned above), as these are very different from the discipline courses taught in the past.  The third step is to convince students who are experiencing a different kind of instruction and assessment that these changes are for their benefit.  Needless to say, none of these steps is easy.

I am proud of what we achieved in our education reforms, but I am not naïve enough to think that these reforms are easy to maintain.  Given the different education philosophies outlined earlier, reforms can be stopped or even reversed.  This will be a constant and long struggle.  I hope that more scientists will think about how we educate our next generation.   They are the people who will keep science alive.

Acknowledgements

I thank my HKU colleagues who helped me implemented the curriculum reform at HKU, in particular N.K. Tsing and Eddy K.F. Lam who were instrumental in the design of the Science Foundation courses.  I am grateful to HsingChi von Bergmann and Carl Pennypacker of the Global Science Education Network for their advice on the reform.

References

[1]Heffron, J.M. The knowledge most worth having: Otis W. Caldwell (1869-1947) and the rise of the general science course, *Science and Education*, 4, 227-252 (1995).

[2]National Commission on Excellence in Education, A nation at risk: the imperative for education reform (1983).

[3]Jaffee, D.  The general education initiatives in Hong Kong: organized contradictions and emerging tensions, *High Education*, Volume 64, Issue 2, pp 193–206 (2012).

[4]Kwok, S.  *Our Place in the Universe: understanding fundamental astronomy from ancient discoveries*, Springer (2017).

[5]Adams, J.V. Student evaluations: the ratings game, Inquiry, 1, 10-16 (1997).

[6]Deslauriers, L., Schelew, E., and Wieman, C.  Improved Learning in a Large-Enrollment Physics Class. Science, 332, 862-864 (2011).